\documentclass[preprint,authoryear,12pt]{elsarticle}

\usepackage{epsfig}

\usepackage{amssymb}
\usepackage[
a4paper=true,%
breaklinks=true,%
colorlinks=true,%
pdfauthor={First Author et al.},%
pdftitle={Template for manuscripts in Advances in Space Research}%
]{hyperref}

\def\apj{Astroph. J.\/}
\def\apjl{Astroph. J. Lett.\/}
\def\aj{Astron. J.\/}

\def\aap{Astron. Astroph. \/}

\journal{Advances in Space Research}

\begin{document}

\begin{frontmatter}



\title{Exploring Low Luminosity Quasar Diversity at $z \approx$ 2.5 with the Gran Telescopio Canarias\tnoteref{footnote1}}
\tnotetext[footnote1]{Contributed talk presented on May 14, 2013 at the 9th SCSLSA held in Banja Kovilia\v{c}a, Serbia.}


\author{Jack  W. Sulentic\corref{cor} and Ascensi\'{on} del Olmo}
\address{Instituto de Astrof{\'\i}sica de Andaluc{\'\i}a (CSIC), 18008 Granada, Espa\~na}
\ead{sulentic@iaa.es and chony@iaa.es}

\author{Paola Marziani}
\address{INAF, Osservatorio Astronomico di Padova, 35122 Padova, Italia}
\ead{paola.marziani@oapd.inaf.it}



\begin{abstract}

We present preliminary results from a pencil-beam spectroscopic  survey of 
low-luminosity quasars at $z \approx 2.2-2.5$. Our goal is to compare these sources 
with low redshift analogues of similar luminosity. High s/n and moderate resolution 
spectra were obtained for 15 sources  using the faint object spectrograph  Osiris 
on the 10m Gran Telescopio Canarias. The new data make possible an almost unprecedented  comparison between sources with the same (moderate) luminosity at widely different 
cosmic epochs. Preliminary analysis of our spectra confirms the presence of a 
relatively evolved population of quasars radiating at modest $L/L_\mathrm{Edd}$. 
A notable difference between the low and high $z$\ quasars may involve 
the presence of lower metallicity quasars at high redshift.  
\end{abstract}

\end{frontmatter}

\parindent=0.5 cm


\section{Introduction}

It has been known for decades that quasars show evolution with redshift.
The brightest quasars at high redshift are much more luminous than any 
local quasars. Locally ($z \lesssim $0.5) we find quasars with absolute magnitudes in 
the range $M_\mathrm{B} \approx $  --22 to --25 (we ignore here AGN below --22).  At redshift $z \approx$ 2.5 we 
find sources in the range $M_{r}$ = --25 to --29. Quasars in this luminosity  range do not 
exist locally.  At redshift $z \approx 4$\ we observe M$_\mathrm{B}$ = --27 to --30 superluminous 
quasars that are more than 2dex above any local sources. Figure 1 illustrates this 
situation quite  well by plotting the distribution of a representative  SDSS quasar 
subsample  \citep{schneideretal10} in the $z$ -- $B$ absolute magnitude  plane. If analogues  to local  quasars exist at all redshifts then the lower right quadrant of the plot will be  well populated when they are found: they are generally fainter than $m_\mathrm{B} \approx$ 22 and are  therefore not so easy to find. 

When we compare in some detail quasars at significantly different redshifts  we find it 
difficult to compare objects with similar luminosities.  We tend to compare the brightest  
quasars at each redshift because we can obtain better data with smaller telescopes  
and less observing time. This limitation is quite important  since luminosity is likely to 
play an important role in quasar physics and spectroscopy provides the most powerful 
clues about quasar kinematics, structure  and evolution. Obtaining spectra for  $z \geq $ 2.2-2.5 
luminosity analogues of nearby quasars/Seyfert galaxies involves spectroscopy 
of faint sources $m_\mathrm{B}$ = 21 -- 23.   It is one thing to obtain spectra in order to 
confirm that faint sources are quasars and quite another to obtain spectra with s/n and resolution enabling more detailed study. 
 
One might ask if high redshift analogues of local low luminosity quasars even exist? 
They are  assumed to be plentiful, perhaps even more numerous that at low redshift. 
Do they show similar emission line properties and derived  black holes masses and Eddington ratios? The few studies of low $L$ quasars at high $z$\ have focussed more on estimating the space density of these sources \citep{glikmanetal11,ikedaetal12}. We focus instead on trying to answer the questions raised above.The faint quasars of interest are largely absent in Figure 1 because they were not sampled by SDSS. They are more likely to be discovered serendipitously 
in deep radio, X-ray or optical 
pencil beam surveys. SDSS surveys quasars largely brighter than $m_\mathrm{B}$ = 20 -- 21 at redshifts above $z \approx$ 1.0.  If the high redshift analogues of low luminosity/redshift  quasars exist in abundance are they spectroscopically different? Using existing spectra of such high $z$ and low $L$\ quasars we 
can do little more than confirm that they are quasars.

One of the many technological changes that has taken place over the 50 years of
quasar science  involves the advent of large telescopes equipped with CCD
imaging spectrographs. We are now entering the era when we can effectively
compare quasars of similar luminosities at high and low redshift. We
report here on such a pilot survey of $z \approx $2.5 quasars with luminosities
similar to the brightest ones observed locally.

\begin{figure}

\begin{center}
\includegraphics*[width=13.25cm,angle=0]{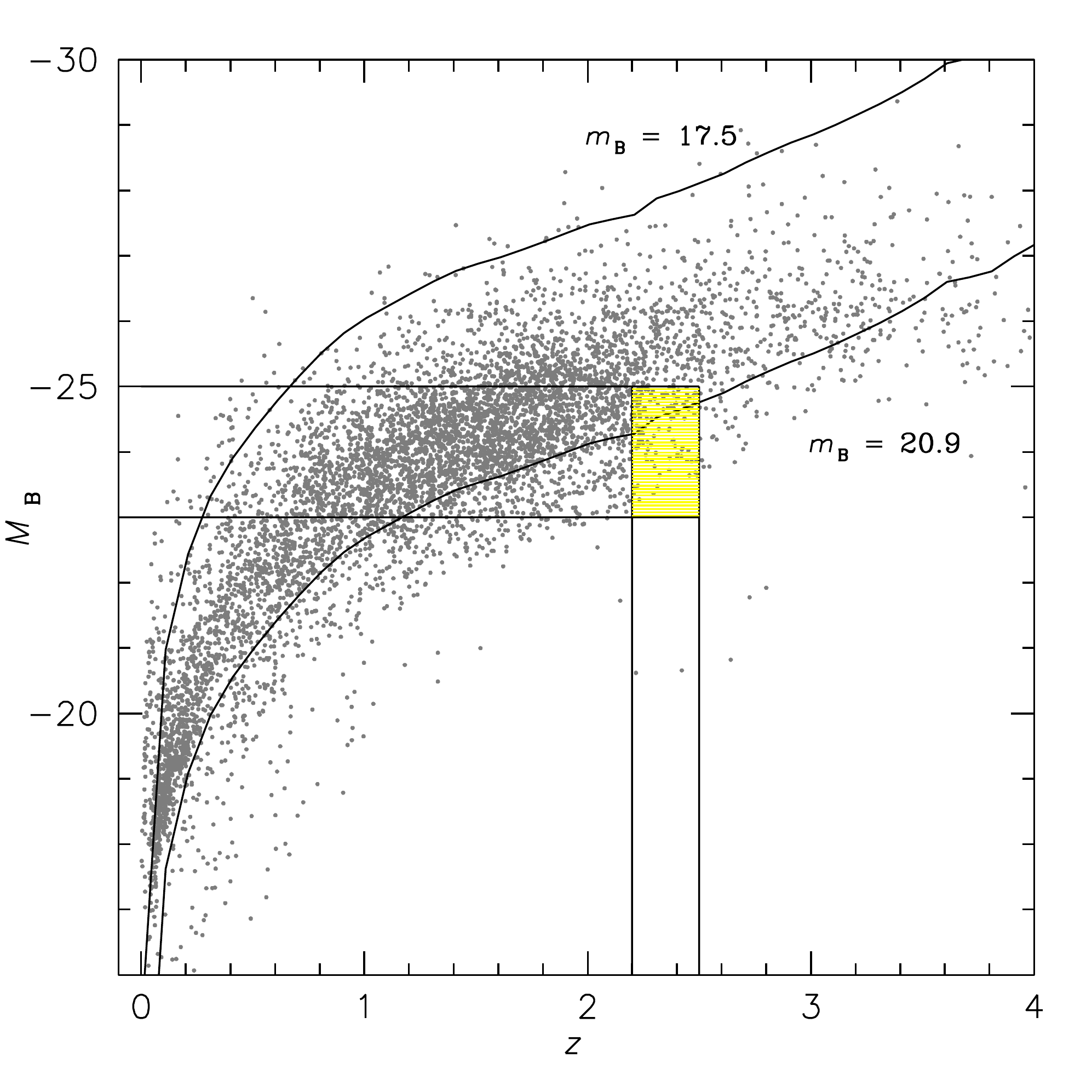}
\end{center}
\caption{Absolute magnitude of quasars  vs. redshift for  the SDSS-based  catalog of \cite{schneideretal10}. Curves are computed for two limiting (K-corrected) apparent magnitudes 
$m_\mathrm{B} = 17.5$  and $m_\mathrm{B} = 20.9$. The shaded box identifies  an example of volume 
limited between $2.2 \le z \le 2.5$.  Luminosity sampling decreases strongly with redshift. Only the high  
end of the quasar luminosity function can be sampled at high $z$ even with relatively deep surveys.  \label{fig:mabsz}}
\end{figure}

\section{A Pencil-beam survey at $z \approx$2.5: Sample selection}

Figure \ref{fig:mabsz} shows that we are sampling quasars with different luminosities at every $z$.  If quasars 
are the same at all redshifts then we could study their optical luminosity function  by combining spectra 
for sources over a wide redshift range. High redshift sources would provide the bright end and local quasars
the faint end. This would be a dangerous thing to do given the degree of luminosity evolution implied by
Figure 1. Our own comparison of luminosity binned composite spectra also show differences \citep{marzianietal09}
that argue against the above procedure. Using the 10m GTC equipped with Osiris we can 
obtain  S/N $\approx$ 20 spectra of quasars in the $m_\mathrm{B}$ =21 -- 22 magnitude range with exposures of about 40 minutes. 
At this redshift we cover the wavelength range from Lyman$\alpha$ to C{\sc iii}]$\lambda$1909 (1000 -- 2500 \AA) providing many spectroscopic diagnostics. Our GTC spectra will be similar to, or  better than, HST FOS archival UV spectra for low redshift analogues \citep{bachevetal04,sulenticetal07}.  Both the high and low z sources show bolometric luminosities near $\log L_\mathrm{bol} \sim $46 the most luminous observed at low $z$.
We had hoped to explore less luminous quasars (e.g. the brightest Seyfert 1´s) but  required exposure times were too high. 

\begin{figure}
\begin{center}
\includegraphics*[width=6.25cm,angle=0]{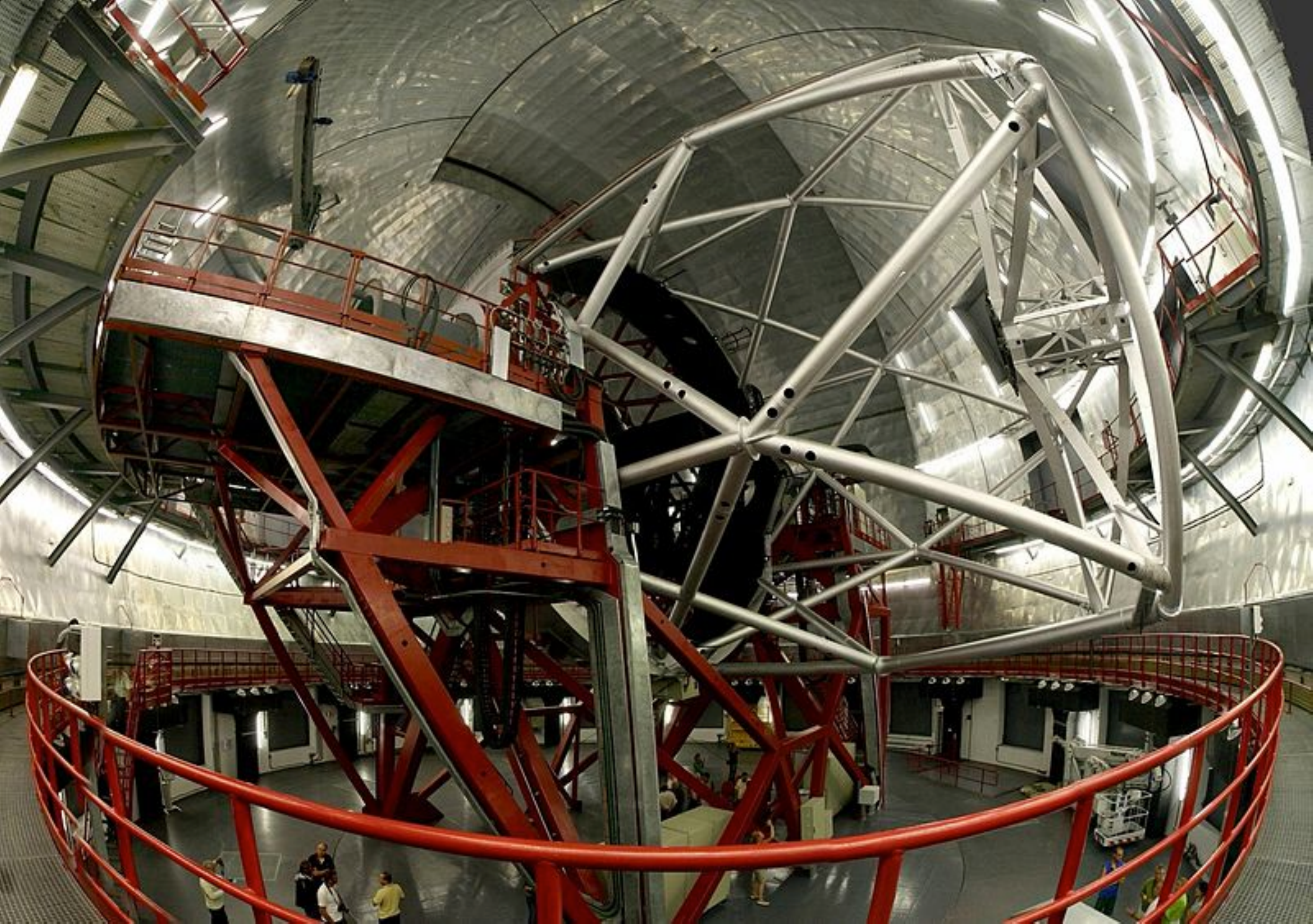}
\includegraphics*[width=6.60cm,angle=0]{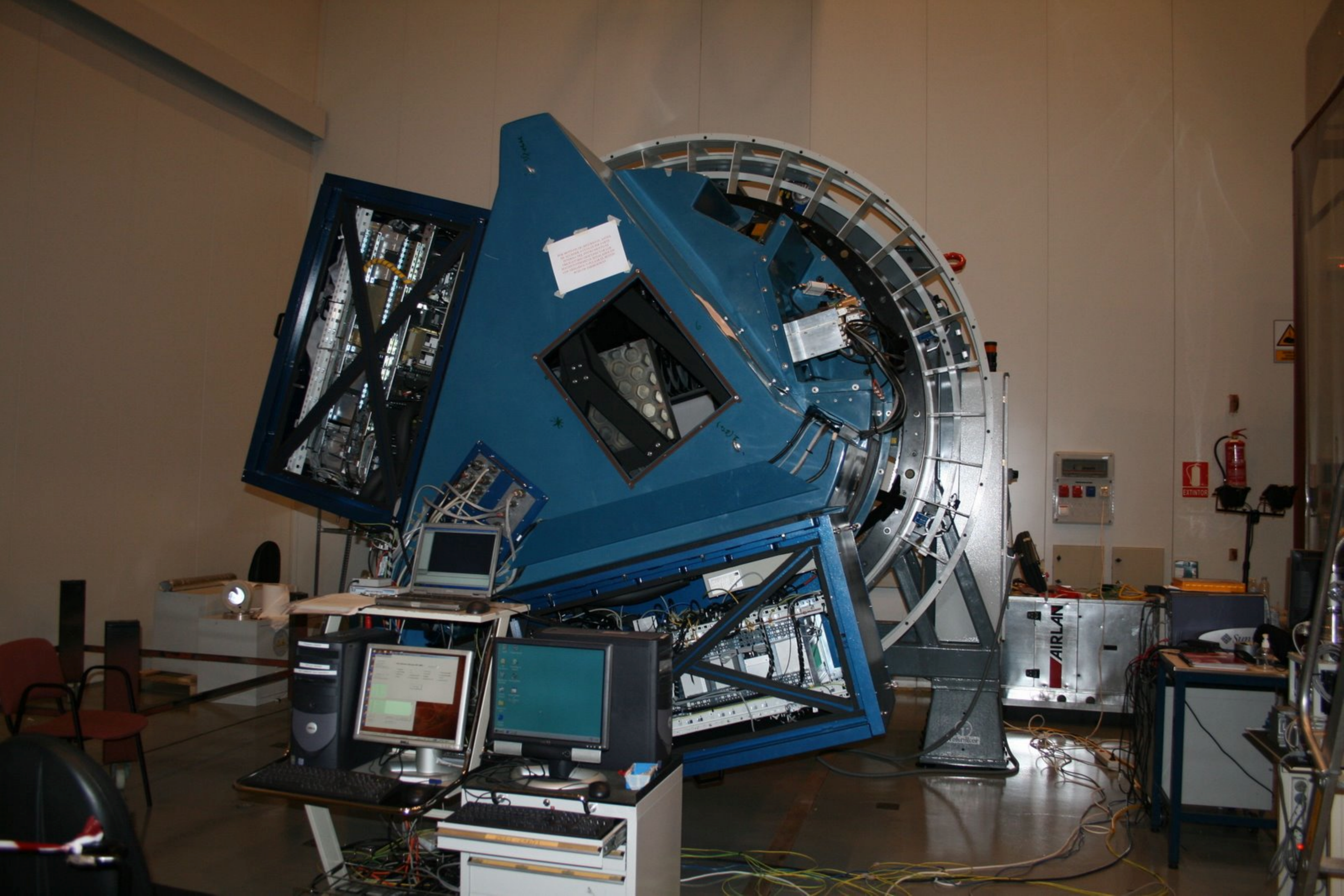}
\end{center}
\caption{The Gran Telescopio Canarias (left) and the Osiris faint object spectrographic camera mounted at its Nasmith focus  (right). \label{fig:gtc}}
\end{figure}

\begin{figure}
\begin{center}
\includegraphics*[width=6.75cm,angle=0]{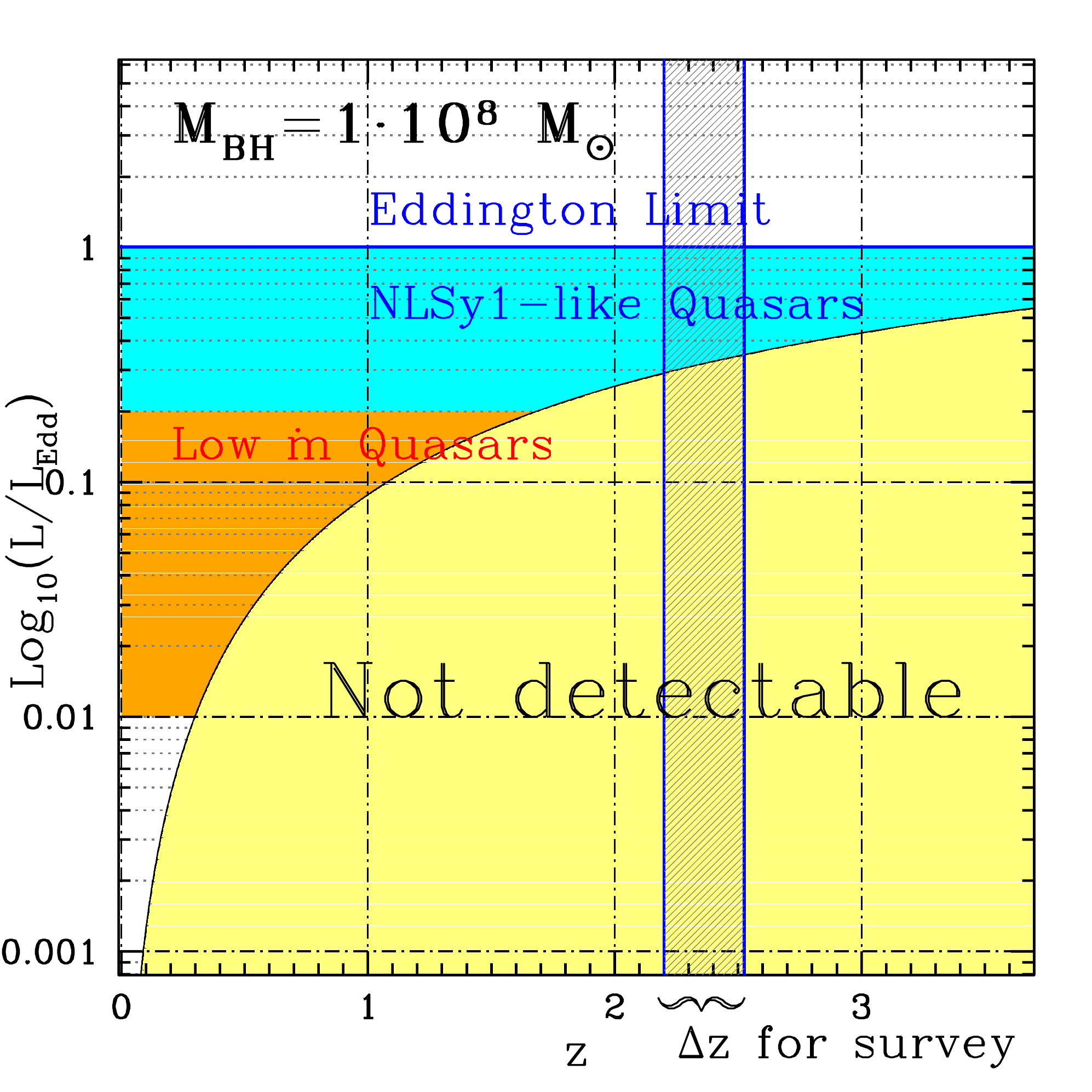}
\includegraphics*[width=6.75cm,angle=0]{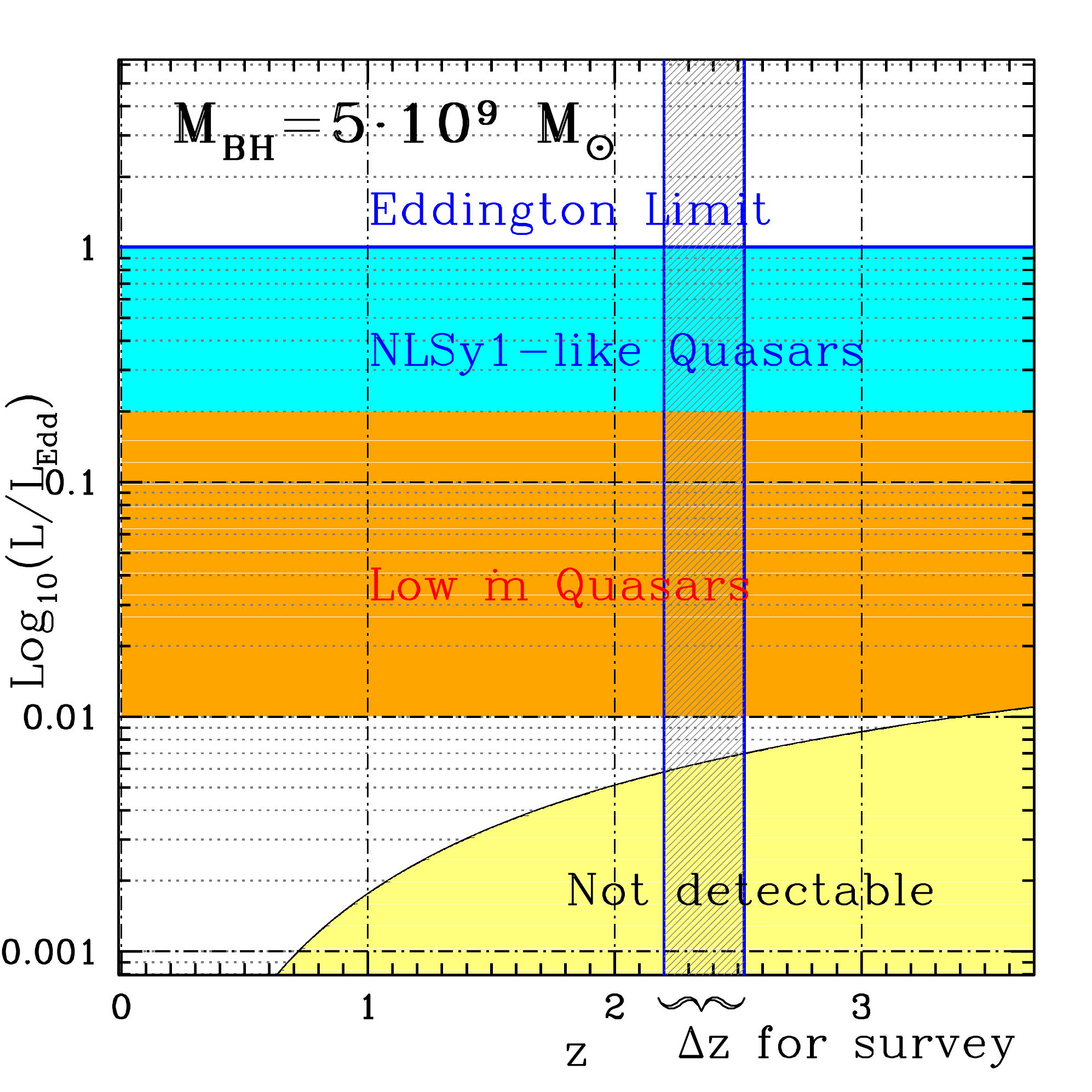}
\end{center}
\caption{Eddington ratio as a function of redshift, with the area of undetectable sources below a limiting magnitude $m_r \approx 21.5$) in yellow. The redshift range of the  quasar survey is identified by the dashed strip. See text for further details. \label{fig:eb}}
\end{figure}

Steps to find low luminosity quasars  at $z \approx 2.5.$ involve a search for faint identified quasars  in 
the V\'eron-Cetty \& V\'eron catalog  \citep[e.g.,][]{veroncettyveron10}:  since we do not carry out  a new survey. Many of our adopted targets  are SDSS sources however all  but one are listed as stars. This study makes use of the Gran Telescopio Canarias (GTC) equipped with an efficient faint object spectrographic camera (Osiris at the Nasmyth focus of GTC). Figure \ref{fig:gtc} shows the 10m telescope  and Osiris configuration attached to the Nasmyth focus.  A sample of sources was selected in narrow redshift  ($2.3 \lesssim z \lesssim 2.5$) and apparent magnitude 
(20 -- 22) ranges ($-25 \lesssim M_{V} \lesssim -23$). This will ensure that we are observing low-luminosity quasars about 1dex more luminous than  the nominal limit for Seyfert nuclei.  As expected many have names reflecting discovery in X-ray surveys or deep optical searches -- so we cannot claim our sample is unbiased.  A strong evolution of Eddington ratio is expected \citep{padovani89,cavalierevittorini00}. The flux limits associated with optical surveys cause an ``Eddington bias'' as shown in Figure \ref{fig:eb}. The mass-dependent loss of low Eddington ratio sources  is one potential source of bias.  

\section{Observations}

\begin{figure}
\begin{center}
\includegraphics*[width=13.25cm,angle=0]{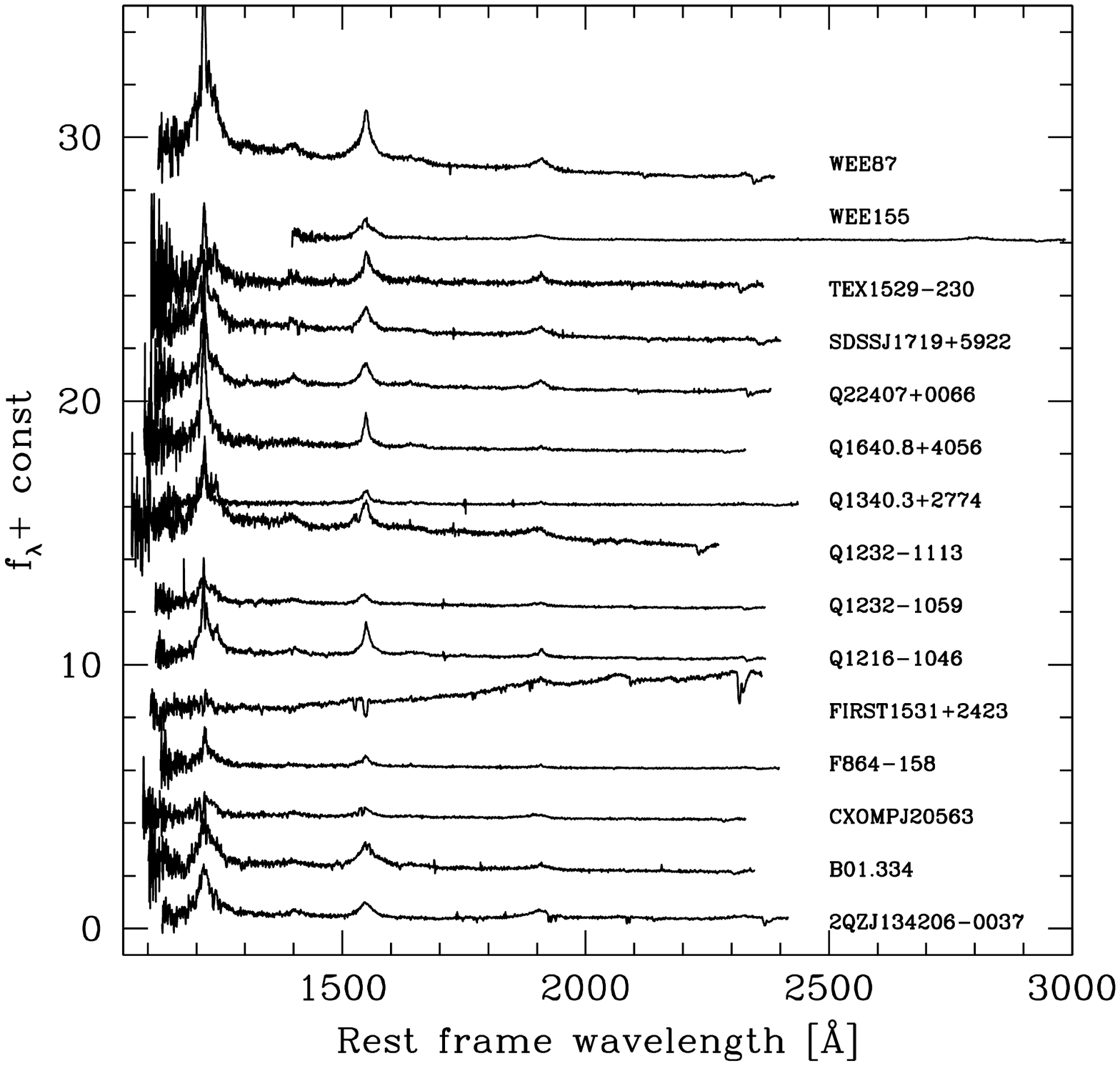}
\end{center}
\caption{The first 15 GTC quasar spectra after redshift correction. Abscissa is the rest frame wavelength in \AA, ordinate is specific flux in units of 10$^{-15}$ erg s$^{-1}$ cm$^{-2}$ \AA$^{-1}$. Spectra have been vertically  displaced by adding  steps of $\Delta f_{\lambda} = 2$  from bottom to top.  \label{fig:overview} }
\end{figure}

Figure \ref{fig:overview} shows a montage of the deredshifted GTC spectra obtained for 15 quasars. One source (Wee 155) 
turned out to have an incorrect redshift ($z\approx$1.59) based on the low S/N discovery spectrum. In addition 
one source is a heavily reddened (radio-loud) quasar FIRST J15318+2423  with redshift $z\approx$2.28. The remaining 13  sources are confirmed as low luminosity unreddened quasars (3 radio-loud) near $z \approx$2.5. Spectra show  S/N near 15 -- 20.   The optical spectra covering Ly$\alpha$ -- C{\sc iii}]$\lambda$1909 in the rest frame  can be  compared with low $z$\ UV spectra from the HST archive \citep{bachevetal04,sulenticetal07}.
We confirm the expectation that high redshift analogues of typical low $z$\ quasars exist. The most recent 
 V\'eron-Cetty \& V\'eron catalog has added many more such quasars in this luminosity range. Our sample clusters near 
log L$_\mathrm{bol} \sim $46.0 with final redshift range from  $z = $ 2.21 -- 2.40.

\section{Results}
There are likely  significant numbers of quasars at $z = 2.5$ with luminosities (log L$_\mathrm{bol} \sim10 ^{46}$) similar to local quasars. Their spectra appear generally similar to low $z$\ analogues.Our previously defined 4D Eigenvector formalism \citep{sulenticetal00b,sulenticetal07} identified two quasar  populations based on optical, UV and X-ray measures. We made a first attempt at identifying Pop. A and B sources in the GTC sample using previously established criteria including:  presence/absence of a C{\sc iv}1549 blueshift/asymmetry, EW C{\sc iv} (low for Pop. A and high for Pop. B), low/high FWHM C{\sc iv} and line ratios in the C{\sc iii}]$\lambda$1909 blend \citep{bachevetal04,sulenticetal07}. On this basis high and low accretors ($\propto$ Eddington ratio) could be tentatively identified. Eddington ratio appears to be the principal driver of 4DE1 differences \citep{marzianietal01}.  We find 7 Pop. A and 6 Pop. B sources (omitting the lower z and reddened quasars) essentially the same division as observed locally. So far the low $L$\ quasar population at $z \approx$ 2.5 appears similar to the low $z$ one. The Pop. A  -- B distinction is preserved. 

\begin{figure}
\begin{center}
\includegraphics*[width=11.25cm,angle=0]{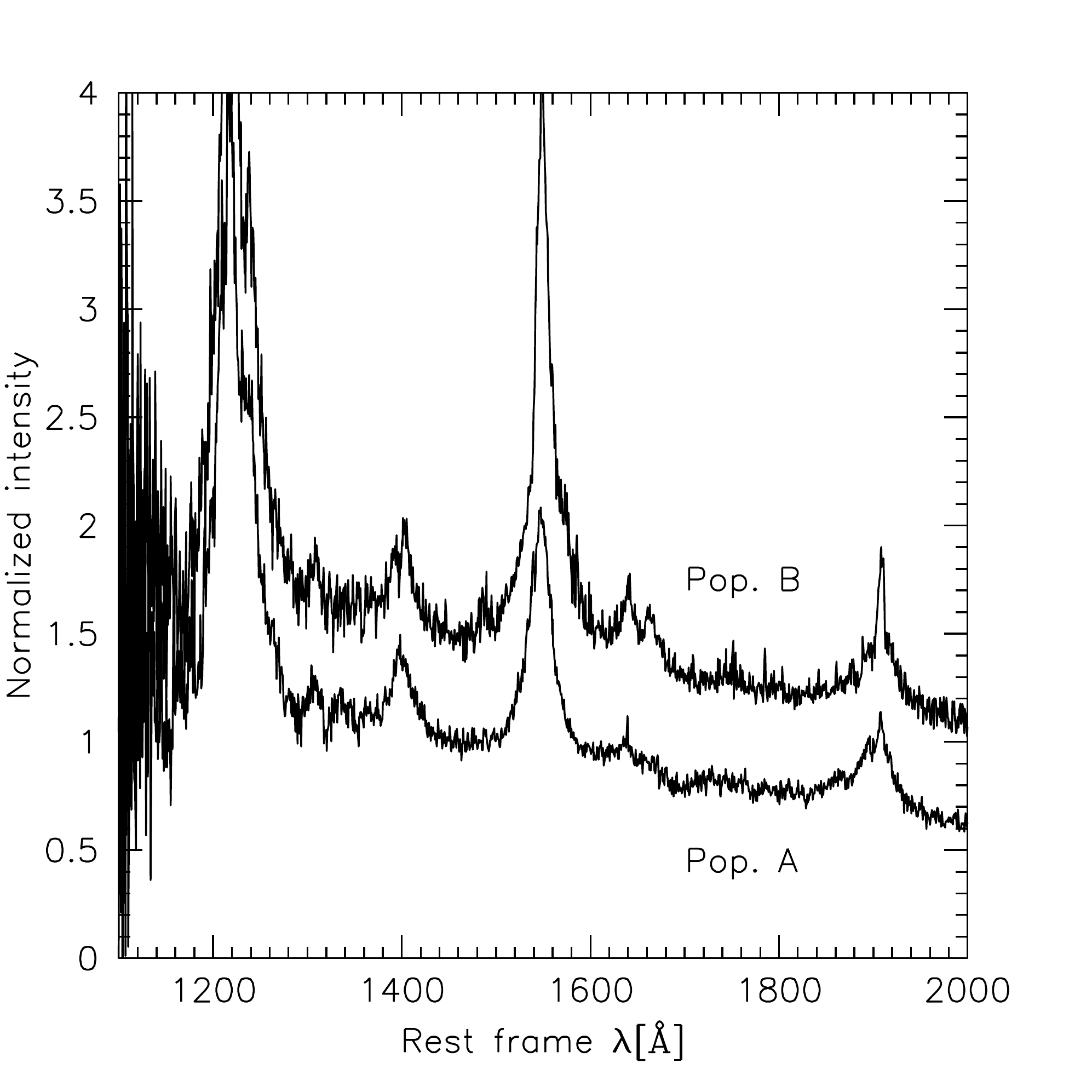}
\end{center}
\caption{Composite spectra obtained computing the median of spectra for Pop. A and B sources separately. \label{fig:popab} }
\end{figure}

\begin{figure}
\begin{center}
\includegraphics*[width=6.5cm,angle=0]{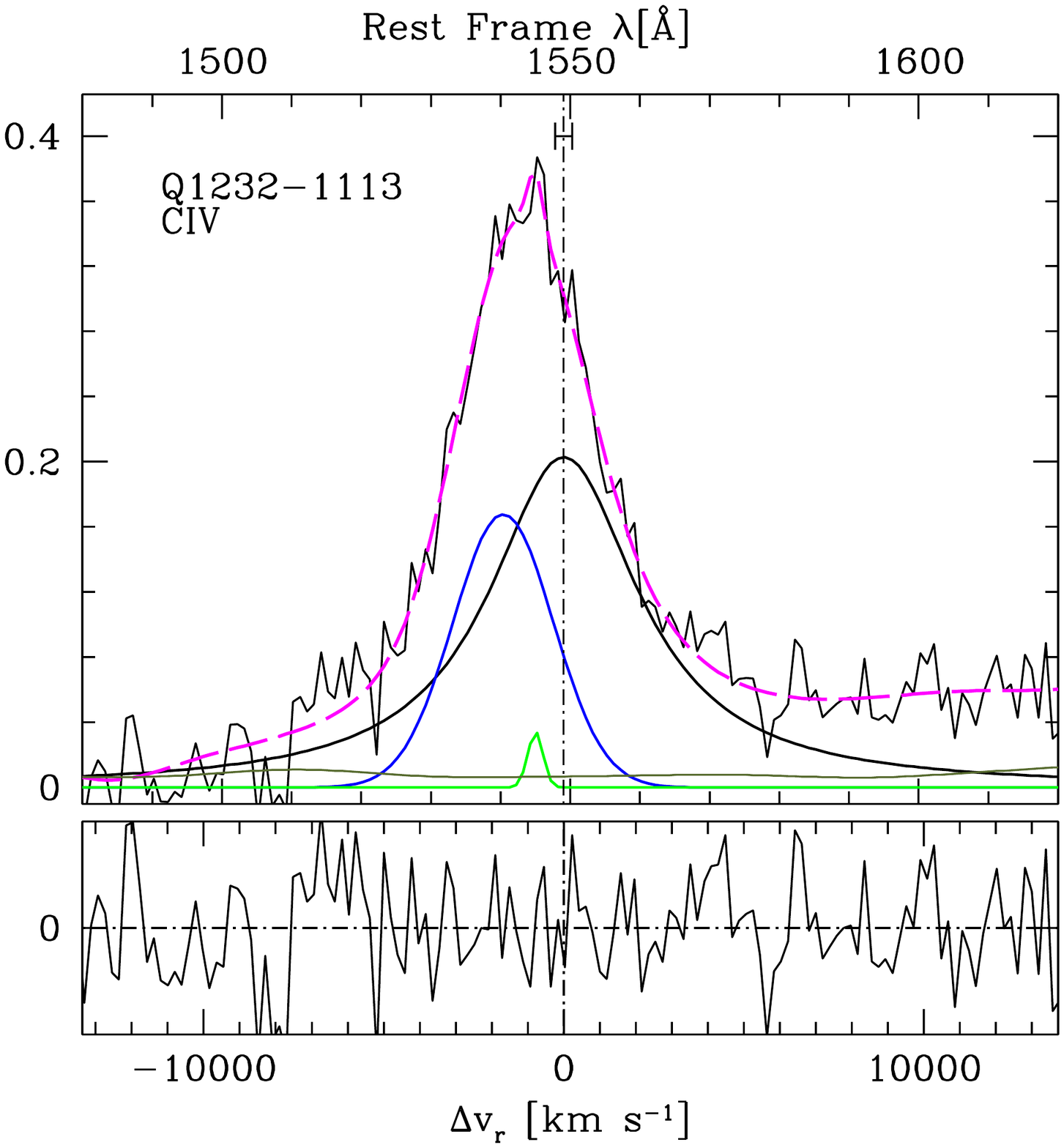}
\includegraphics*[width=6.5cm,angle=0]{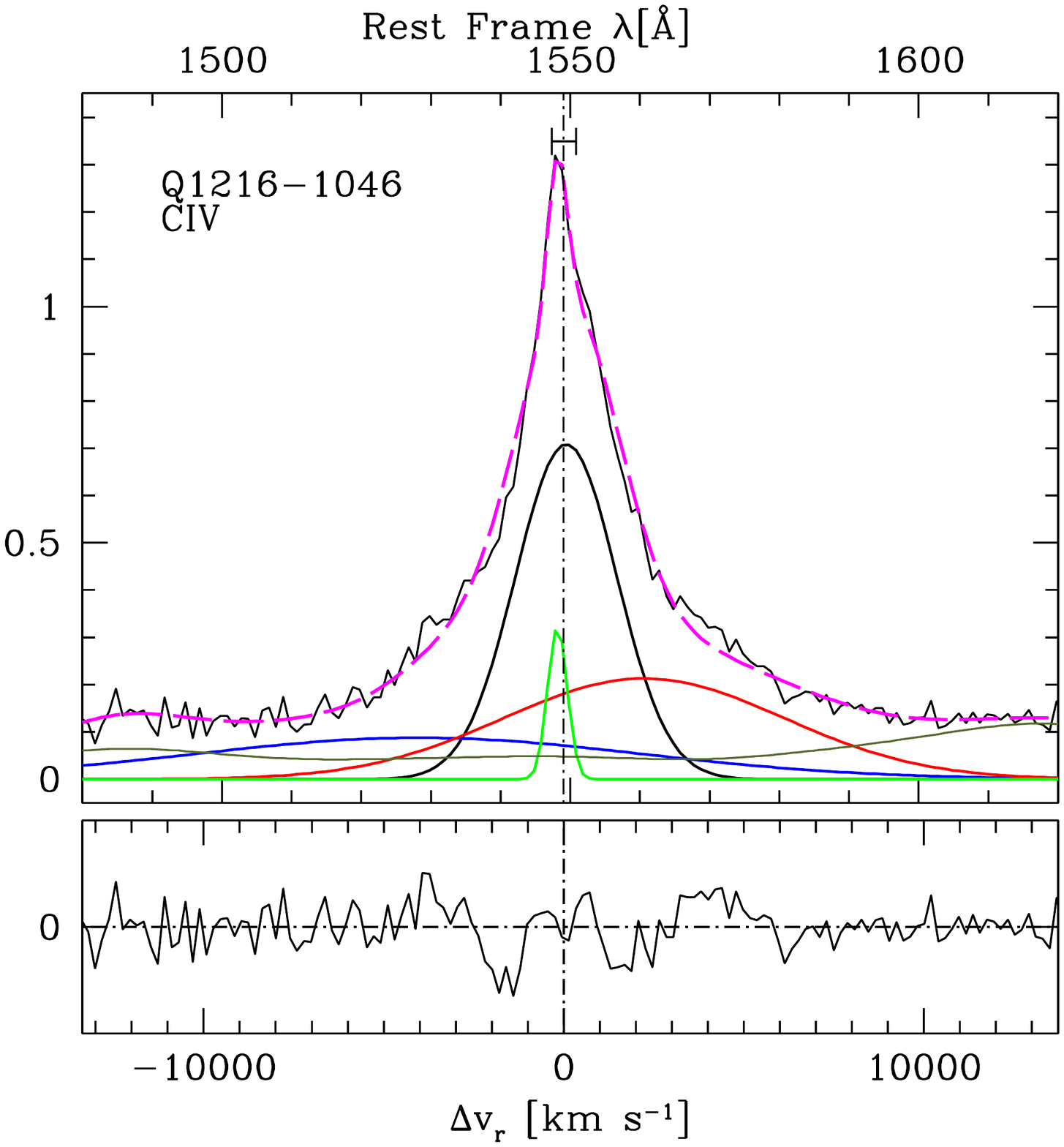}
\end{center}
\caption{Preliminary analysis of  two C{\sc iv}$\lambda$1549 emission line profiles. Source Q1232-1113 is fairly well representative of Pop. A with a large profile shift ascribed to a prominent blue shifted component (blue line).  Source Q1216-1046 is a typical example of Pop. B sources in which the C{\sc iv}$\lambda$1549 line shows large EW ($\approx$100\AA), and no or undetectable blueshift. The redward extension	 of the C{\sc iv}$\lambda$1549 profile requires a redshifted very broad component (red line) to minimize residuals. Green line trace the FeII emission that is however always weak in proximity of C{\sc iv}$\lambda$1549. The black line shows the almost unshifted, symmetric broad component.
\label{fig:civ}}
\end{figure}

Figure \ref{fig:popab} shows Pop. A and B composite spectra constructed from the GTC sample. Pop. A high accretors show weak C{\sc iii}]$\lambda$1909 emission, low C{\sc iv} equivalent width and C{\sc iv} profiles that are blueshifted and blue asymmetric. Narrow line emission is often strong in Pop. B spectra. All detected radio louds are Pop. B.sources.  The generally higher ionization Pop. B spectra also show strong He{\sc ii}$\lambda$1640 and [O{\sc iii}]$\lambda$1663. Figure \ref{fig:civ} shows results of IRAF SPECFIT modelling  of C{\sc iv} in the Pop. A and B composites where differences similar to the ones observed at low z can be clearly seen. 
We do not confirm a near ubiquity of blueshifts/asymmetries for Type 1 quasars \citep{richardsetal11}  but rather the same Pop. A/B dichotomy 
seen at low redshift \citep{sulenticetal07} where Pop. B sources  show 
redshifts/asymmetries or a symmetric 
unshifted profile.

\subsection{A population of low accreting quasars at high redshift}

Our first  attempt at estimating black hole masses and resultant Eddington ratios finds no significant difference between population A and B sources. This is likely because we used FWHM C{\sc iv}$\lambda$1549 as a virial estimator. Comparison of FWHM C{\sc iv}$\lambda$1549 and FWHM H$\beta$\ for low redshift quasars reveals why we find no difference. FWHM C{\sc iv}$\lambda$1549 measures for Pop. A sources even  show a different trends relative to FWHM H$\beta$  virtually precluding accurate black holes mass estimates using FWHM C{\sc iv}$\lambda$1549. Measures from the 1909 blend are a more reliable way to make such estimations. FWHM (of the intermediate ionization lines) Al{\sc iii}$\lambda$1860 (single component of the doublet) and Si{\sc iii}]$\lambda$1892 are effective surrogates for FWHM H$\beta$\ \citep[][and Alenka's Negrete communication in these proceedings]{negreteetal13}.  If the broad component of these lines is used, Eddington ratio is systematically higher for Pop. A sources as found at low redshift. The observed range  is $-1.5 \lesssim L/L_\mathrm{Edd} \lesssim -0.7$\ and indicates moderately accreting sources. Black hole masses range between $10^8$\ and $10^9$\ solar masses.  

One advantage of comparisons using UV spectra at high and low redshifts involves the numerous line ratios that can serve as metallicity indicators. Most results of the past decade point toward solar and supersolar metallicity even at very high redshifts \citep{kurketal07,willottetal10}. Of course these studies focussed on extremely  luminous sources. Our low $L$ sample uses spectra with s/n high enough to permit reasonable measures of metallicity indicators (NV$\lambda$1240/C{\sc iv}, Si{\sc iv}+O{\sc iv}]/C{\sc iv}). We tentatively identify at least three quasars (F864-158, Q 1340+27 and Q 1640+40) that  appear to show  subsolar or solar metallicities following the normalization of \citet{nagaoetal06}. The detection of  those object hints  at a diversity  in chemical abundances consistent with the one found at low-$z$ from an analysis of PG quasars \citep{shinetal13}. 

 \section{Discussion: are there real evolutionary effects?}

Small samples are prone to statistical fluctuation.  Considering that the Baldwin effect is a rather weak correlation, there is no point in claiming a detection (or a non detection) of a Baldwin effect unless the sample  exceeds $\sim 100$\ quasars \citep{sulenticetal00a}.  In addition, the equivalent width dispersion is  not due to a luminosity correlation since  we are studying a sample with essentially fixed luminosity.  Our first estimates of EW C{\sc iv} range from 40 -- 150\AA\  which reflects the diversity found at low z \citep{bachevetal04} where mean EW C{\sc iv} values of 30 and $\sim$100 were found for extreme Pop. A and B sources respectively.  

In summary, the detection of sources radiating at modest Eddington ratios, the high fraction of Pop. B sources that are not wind dominated, the modest Eddington ratios as well as metallicity properties suggest that the diversity in quasar properties is consistent with the diversity found in the local Universe. These findings do not bode well with a strong evolution between the present cosmic epoch and an earlier epoch $\approx 3 $ Gyr after the Big Bang. 

A full elaboration of this comparison between low and high redshift quasars is under preparation for an increased high-$z$\ sample of 22 sources.

\bibliographystyle{model1b-num-names} 

\begin{thebibliography}{17}
\expandafter\ifx\csname natexlab\endcsname\relax\def\natexlab#1{#1}\fi
\providecommand{\bibinfo}[2]{#2}
\ifx\xfnm\relax \def\xfnm[#1]{\unskip,\space#1}\fi
\bibitem[{{Bachev} et~al.(2004){Bachev}, {Marziani}, {Sulentic}, {Zamanov},
  {Calvani} and {Dultzin-Hacyan}}]{bachevetal04}
\bibinfo{author}{R.~{Bachev}}, \bibinfo{author}{P.~{Marziani}},
  \bibinfo{author}{J.W. {Sulentic}}, \bibinfo{author}{R.~{Zamanov}},
  \bibinfo{author}{M.~{Calvani}}, \bibinfo{author}{D.~{Dultzin-Hacyan}},
  \bibinfo{title}{{Average Ultraviolet Quasar Spectra in the Context of
  Eigenvector 1: A Baldwin Effect Governed by the Eddington Ratio?}},
  \bibinfo{journal}{Astroph. J.\/} \bibinfo{volume}{617} (\bibinfo{year}{2004})
  \bibinfo{pages}{171--183}.
\bibitem[{{Cavaliere} and {Vittorini}(2000)}]{cavalierevittorini00}
\bibinfo{author}{A.~{Cavaliere}}, \bibinfo{author}{V.~{Vittorini}},
  \bibinfo{title}{{The Fall of the Quasar Population}}, \bibinfo{journal}{\apj}
  \bibinfo{volume}{543} (\bibinfo{year}{2000}) \bibinfo{pages}{599--610}.
\bibitem[{{Glikman} et~al.(2011){Glikman}, {Djorgovski}, {Stern}, {Dey},
  {Jannuzi} and {Lee}}]{glikmanetal11}
\bibinfo{author}{E.~{Glikman}}, \bibinfo{author}{S.G. {Djorgovski}},
  \bibinfo{author}{D.~{Stern}}, \bibinfo{author}{A.~{Dey}},
  \bibinfo{author}{B.T. {Jannuzi}}, \bibinfo{author}{K.S. {Lee}},
  \bibinfo{title}{{The Faint End of the Quasar Luminosity Function at z \~{} 4:
  Implications for Ionization of the Intergalactic Medium and Cosmic
  Downsizing}}, \bibinfo{journal}{\apjl} \bibinfo{volume}{728}
  (\bibinfo{year}{2011}) \bibinfo{pages}{L26}.
\bibitem[{{Ikeda} et~al.(2012){Ikeda}, {Nagao}, {Matsuoka}, {Taniguchi},
  {Shioya}, {Kajisawa}, {Enoki}, {Capak}, {Civano}, {Koekemoer}, {Masters},
  {Morokuma}, {Salvato}, {Schinnerer} and {Scoville}}]{ikedaetal12}
\bibinfo{author}{H.~{Ikeda}}, \bibinfo{author}{T.~{Nagao}},
  \bibinfo{author}{et {al.}}, \bibinfo{title}{{Constraints on the Faint
  End of the Quasar Luminosity Function at z \~{} 5 in the COSMOS Field}},
  \bibinfo{journal}{\apj} \bibinfo{volume}{756} (\bibinfo{year}{2012})
  \bibinfo{pages}{160}.
\bibitem[{{Kurk} et~al.(2007){Kurk}, {Walter}, {Fan}, {Jiang}, {Riechers},
  {Rix}, {Pentericci}, {Strauss}, {Carilli} and {Wagner}}]{kurketal07}
\bibinfo{author}{J.D. {Kurk}}, \bibinfo{author}{F.~{Walter}},
  \bibinfo{author}{X.~{Fan}}, \bibinfo{author}{L.~{Jiang}},
  \bibinfo{author}{D.A. {Riechers}}, \bibinfo{author}{H.W. {Rix}},
  \bibinfo{author}{L.~{Pentericci}}, \bibinfo{author}{M.A. {Strauss}},
  \bibinfo{author}{C.~{Carilli}}, \bibinfo{author}{S.~{Wagner}},
  \bibinfo{title}{{Black Hole Masses and Enrichment of z \~{} 6 SDSS Quasars}},
  \bibinfo{journal}{\apj} \bibinfo{volume}{669} (\bibinfo{year}{2007})
  \bibinfo{pages}{32--44}.
\bibitem[{{Marziani} et~al.(2009){Marziani}, {Sulentic}, {Stirpe}, {Zamfir} and
  {Calvani}}]{marzianietal09}
\bibinfo{author}{P.~{Marziani}}, \bibinfo{author}{J.W. {Sulentic}},
  \bibinfo{author}{G.M. {Stirpe}}, \bibinfo{author}{S.~{Zamfir}},
  \bibinfo{author}{M.~{Calvani}}, \bibinfo{title}{{VLT/ISAAC spectra of the
  H{$\beta$} region in intermediate-redshift quasars. III. H{$\beta$}
  broad-line profile analysis and inferences about BLR structure}},
  \bibinfo{journal}{Astron. Astroph.} \bibinfo{volume}{495} (\bibinfo{year}{2009})
  \bibinfo{pages}{83--112}.
\bibitem[{{Marziani} et~al.(2001){Marziani}, {Sulentic}, {Zwitter},
  {Dultzin-Hacyan} and {Calvani}}]{marzianietal01}
\bibinfo{author}{P.~{Marziani}}, \bibinfo{author}{J.W. {Sulentic}},
  \bibinfo{author}{T.~{Zwitter}}, \bibinfo{author}{D.~{Dultzin-Hacyan}},
  \bibinfo{author}{M.~{Calvani}}, \bibinfo{title}{{Searching for the Physical
  Drivers of the Eigenvector 1 Correlation Space}}, \bibinfo{journal}{ApJ}
  \bibinfo{volume}{558} (\bibinfo{year}{2001}) \bibinfo{pages}{553--560}.
\bibitem[{{Nagao} et~al.(2006){Nagao}, {Marconi} and {Maiolino}}]{nagaoetal06}
\bibinfo{author}{T.~{Nagao}}, \bibinfo{author}{A.~{Marconi}},
  \bibinfo{author}{R.~{Maiolino}}, \bibinfo{title}{{The evolution of the
  broad-line region among SDSS quasars}}, \bibinfo{journal}{A\&Ap}
  \bibinfo{volume}{447} (\bibinfo{year}{2006}) \bibinfo{pages}{157--172}.
\bibitem[{{Negrete} et~al.(2013){Negrete}, {Dultzin}, {Marziani} and
  {Sulentic}}]{negreteetal13}
\bibinfo{author}{C.A. {Negrete}}, \bibinfo{author}{D.~{Dultzin}},
  \bibinfo{author}{P.~{Marziani}}, \bibinfo{author}{J.w. {Sulentic}},
  \bibinfo{title}{{Reverberation and photoionization estimates of the Broad
  Line Region Radius in Low-z Quasars}}, \bibinfo{journal}{Astroph. J.}
  \bibinfo{volume}{771} (\bibinfo{year}{2013})
  \bibinfo{pages}{31, 18pp}.
  (\bibinfo{year}{2013}).
\bibitem[{{Padovani}(1989)}]{padovani89}
\bibinfo{author}{P.~{Padovani}}, \bibinfo{title}{{The evolution of the
  Eddington ratio for active galactic nuclei}}, \bibinfo{journal}{\aap}
  \bibinfo{volume}{209} (\bibinfo{year}{1989}) \bibinfo{pages}{27--45}.
\bibitem[{{Richards} et~al.(2011){Richards}, {Kruczek}, {Gallagher}, {Hall},
  {Hewett}, {Leighly}, {Deo}, {Kratzer} and {Shen}}]{richardsetal11}
\bibinfo{author}{G.T. {Richards}}, \bibinfo{author}{N.E. {Kruczek}},
  \bibinfo{author}{S.C. {Gallagher}}, \bibinfo{author}{P.B. {Hall}},
  \bibinfo{author}{P.C. {Hewett}}, \bibinfo{author}{K.M. {Leighly}},
  \bibinfo{author}{R.P. {Deo}}, \bibinfo{author}{R.M. {Kratzer}},
  \bibinfo{author}{Y.~{Shen}}, \bibinfo{title}{{Unification of Luminous Type 1
  Quasars through C IV Emission}}, \bibinfo{journal}{\aj} \bibinfo{volume}{141}
  (\bibinfo{year}{2011}) \bibinfo{pages}{167, 16pp}.
\bibitem[{{Schneider} et~al.(2010){Schneider}, {Richards}, {Hall}, {Strauss},
  {Anderson}, {Boroson}, {Ross}, {Shen}, {Brandt}, {Fan}, {Inada}, {Jester},
  {Knapp}, {Krawczyk}, {Thakar}, {Vanden Berk}, {Voges}, {Yanny}, {York},
  {Bahcall}, {Bizyaev}, {Blanton}, {Brewington}, {Brinkmann}, {Eisenstein},
  {Frieman}, {Fukugita}, {Gray}, {Gunn}, {Hibon}, {Ivezi{\'c}}, {Kent}, {Kron},
  {Lee}, {Lupton}, {Malanushenko}, {Malanushenko}, {Oravetz}, {Pan}, {Pier},
  {Price}, {Saxe}, {Schlegel}, {Simmons}, {Snedden}, {SubbaRao}, {Szalay} and
  {Weinberg}}]{schneideretal10}
\bibinfo{author}{D.P. {Schneider}}, \bibinfo{author}{G.T. {Richards}},
  \bibinfo{author}{P.B. {Hall}}, \bibinfo{author}{M.A. {Strauss}},
  \bibinfo{author}{S.F. {Anderson}}, \bibinfo{author}{T.A. {Boroson}},
  \bibinfo{author}{N.P. {Ross}}, \bibinfo{author}{et~{al.}},
  \bibinfo{title}{{The Sloan Digital Sky Survey Quasar Catalog. V. Seventh Data
  Release}}, \bibinfo{journal}{\aj} \bibinfo{volume}{139}
  (\bibinfo{year}{2010}) \bibinfo{pages}{2360--2373}.
\bibitem[{{Shin} et~al.(2013){Shin}, {Woo}, {Nagao} and {Kim}}]{shinetal13}
\bibinfo{author}{J.~{Shin}}, \bibinfo{author}{J.H. {Woo}},
  \bibinfo{author}{T.~{Nagao}}, \bibinfo{author}{S.C. {Kim}},
  \bibinfo{title}{{The Chemical Properties of Low-redshift QSOs}},
  \bibinfo{journal}{\apj} \bibinfo{volume}{763} (\bibinfo{year}{2013})
  \bibinfo{pages}{58}.
\bibitem[{{Sulentic} et~al.(2007){Sulentic}, {Bachev}, {Marziani}, {Negrete}
  and {Dultzin}}]{sulenticetal07}
\bibinfo{author}{J.W. {Sulentic}}, \bibinfo{author}{R.~{Bachev}},
  \bibinfo{author}{P.~{Marziani}}, \bibinfo{author}{C.A. {Negrete}},
  \bibinfo{author}{D.~{Dultzin}}, \bibinfo{title}{{C IV {$\lambda$}1549 as an
  Eigenvector 1 Parameter for Active Galactic Nuclei}}, \bibinfo{journal}{Astroph. J.\/}
  \bibinfo{volume}{666} (\bibinfo{year}{2007}) \bibinfo{pages}{757--777}.
\bibitem[{{Sulentic} et~al.(2000{\natexlab{a}}){Sulentic}, {Marziani} and
  {Dultzin-Hacyan}}]{sulenticetal00a}
\bibinfo{author}{J.W. {Sulentic}}, \bibinfo{author}{P.~{Marziani}},
  \bibinfo{author}{D.~{Dultzin-Hacyan}}, \bibinfo{title}{{Phenomenology of
  Broad Emission Lines in Active Galactic Nuclei}}, \bibinfo{journal}{ARA\&A}
  \bibinfo{volume}{38} (\bibinfo{year}{2000}{\natexlab{a}})
  \bibinfo{pages}{521--571}.
\bibitem[{{Sulentic} et~al.(2000{\natexlab{b}}){Sulentic}, {Marziani},
  {Zwitter}, {Dultzin-Hacyan} and {Calvani}}]{sulenticetal00b}
\bibinfo{author}{J.W. {Sulentic}}, \bibinfo{author}{P.~{Marziani}},
  \bibinfo{author}{T.~{Zwitter}}, \bibinfo{author}{D.~{Dultzin-Hacyan}},
  \bibinfo{author}{M.~{Calvani}}, \bibinfo{title}{{The Demise of the Classical
  Broad-Line Region in the Luminous Quasar PG 1416-129}},
  \bibinfo{journal}{Astroph. J.\/L} \bibinfo{volume}{545}
  (\bibinfo{year}{2000}{\natexlab{b}}) \bibinfo{pages}{L15--L18}.
\bibitem[{{V{\'e}ron-Cetty} and {V{\'e}ron}(2010)}]{veroncettyveron10}
\bibinfo{author}{M.P. {V{\'e}ron-Cetty}}, \bibinfo{author}{P.~{V{\'e}ron}},
  \bibinfo{title}{{A catalogue of quasars and active nuclei: 13th edition}},
  \bibinfo{journal}{\aap} \bibinfo{volume}{518} (\bibinfo{year}{2010})
  \bibinfo{pages}{A10}.
\bibitem[{{Willott} et~al.(2010){Willott}, {Delorme}, {Reyl{\'e}}, {Albert},
  {Bergeron}, {Crampton}, {Delfosse}, {Forveille}, {Hutchings}, {McLure},
  {Omont} and {Schade}}]{willottetal10}
\bibinfo{author}{C.J. {Willott}}, \bibinfo{author}{P.~{Delorme}},
  \bibinfo{author}{C.~{Reyl{\'e}}}, \bibinfo{author}{L.~{Albert}},
  \bibinfo{author}{J.~{Bergeron}}, \bibinfo{author}{D.~{Crampton}},
  \bibinfo{author}{X.~{Delfosse}}, \bibinfo{author}{T.~{Forveille}},
  \bibinfo{author}{J.B. {Hutchings}}, \bibinfo{author}{R.J. {McLure}},
  \bibinfo{author}{A.~{Omont}}, \bibinfo{author}{D.~{Schade}},
  \bibinfo{title}{{The Canada-France High-z Quasar Survey: Nine New Quasars and
  the Luminosity Function at Redshift 6}}, \bibinfo{journal}{\aj}
  \bibinfo{volume}{139} (\bibinfo{year}{2010}) \bibinfo{pages}{906--918}.

\end{thebibliography}


\end{document}